\documentclass[twocolumn,prl,amsmath,amssymb,showpacs,superscriptaddress]{revtex4-1}
\usepackage{epsf}      
\usepackage{graphicx}
\usepackage{color}
\usepackage{gensymb}
\usepackage{sidecap}

\begin{document}
\title {Structural and magnetic behavior of Cr$_2$Co$_{(1-x)}$Cr$_x$Al inverse Heusler alloys}
\author{Manisha Srivastava}
\affiliation{Department of Physics, Indian Institute of Technology Delhi, Hauz Khas, New Delhi-110016, India}
\author{Guru Dutt Gupt}
\affiliation{Department of Physics, Indian Institute of Technology Delhi, Hauz Khas, New Delhi-110016, India}
\author{Priyanka Nehla}
\affiliation{Department of Physics, Indian Institute of Technology Delhi, Hauz Khas, New Delhi-110016, India}
\author{Anita Dhaka}
\affiliation{Department of Physics, Sri Aurobindo College, University of Delhi, Malviya Nagar, New Delhi-110017, India}
\author{R. S. Dhaka}
\email{rsdhaka@physics.iitd.ac.in}
\affiliation{Department of Physics, Indian Institute of Technology Delhi, Hauz Khas, New Delhi-110016, India}

\date{\today}                                         

\begin{abstract}

We report the structural and magnetic behavior of single phase inverse Heusler alloys Cr$_2$Co$_{(1-x)}$Cr$_x$Al ($x = $ 0, 0.2, 0.4) using x-ray diffraction (XRD), Raman spectroscopy, isothermal magnetization, and magnetic susceptibility measurements. Interestingly, the Rietveld refinement of XRD data with the space group I$\bar{4}m2$ reveal a tetragonal distortion with c/a ratio around 1.38 in these inverse Heusler structures. The bulk compositions have been confirmed by energy dispersive x-ray spectroscopy measurements. The active Raman mode F$_{2g}$ is observed at 320~cm$^{-1}$, which confirms the X-type Heusler structure as the A2 and B2 type structures are known to be not Raman active. The area of F$_{2g}$ mode decreases with Cr concentration, which indicate the origin of this mode due to Co vibrations.  The isothermal magnetization data confirm the magnetic moment close to zero ($\le$0.02 $\mu_B/f.u.$) at $\approx$70~kOe and negligible coercive field suggest the fully compensated ferrimagnetic nature of these samples. The susceptibility behavior indicates irreversibility between zero-field and field-cooled curves and complex magnetic interactions at low temperatures.

\end{abstract}

\maketitle


The full Heusler alloys are $X_2YZ$ type ternary intermetallic compounds having $X$ and $Y$ as transition metals and $Z$ as main group element or an {\it sp}-element \cite{Felser16}. In recent years many Co-based Heusler alloys (HAs) have been studied extensively due to their structural stability and half-metallic ferromagnetic nature as well as they have an advantage of high Curie temperature (T$_{\rm C}$) \cite{NehlaPRB19, NehlaCCMA} and also show the spin gapless semiconducting properties~\cite{RaniPRB19, OuardiPRL13}. In the family of CoCr based full-Heusler alloys, Co$_2$CrAl has particularly attracted huge attention due to its true half-metallic ferromagnetic (HMF) nature with 3~$\mu_{\rm B}$/{\it f.u.} magnetic moment as per the band structure calculations and the Slater-Pauling rule, respectively \cite{KublerPRB07,Wurmehl06}. It is interesting to note here that Luo {\it et al.} studied the effect of Cr substitution on Co site and showed a transition from HMF to HM fully compensated ferrimagnetic state \cite{LuoPB08}. Moreover, the appreciable value of spin polarization of about 68\% was also reported for Cr$_2$CoAl~\cite{Singh14}. This found to be most important material as by changing the Cr concentration, we can tune the magnetic moment from 3~$\mu_{\rm B}$/f.u. to zero without destroying the half-metallicity \cite{LuoPB08}. These compounds show the largest value of spin density among all known ferromagnetic half-metals so far. Interestingly, few Heusler alloys show half-metallic fully compensated ferrimagnetic behavior \cite{GalanakisPRB07}. These materials act as half-metal as well as gapless semiconductors. In 2013, M. Meinert~{\it et al.} theoretically investigated Cr$_2$CoAl Heusler alloys as completely compensated half-metallic ferrimagnets \cite{Meinert13}. The reason for the low moment (0.01$\mu_{\rm B}$/f.u.) is the internal spin compensation of the transition metals placed at different sites (Cr1: 1.36 $\mu_{\rm B}$, Cr2: -1.49 $\mu_{\rm B}$, Co: 0.30 $\mu_{\rm B}$) and the high curie temperature of these materials is because of their strong local moments \cite{Meinert13}. Moreover, there is also possibility of showing low magnetic moments of alloy due to the anti-ferromagnetically orientation of Cr-Cr neighboring elements \cite{Meinert13, Jin_atom_position}. The main advantage of these alloys is that they develop very low stray field or demagnetizing field, due to which they can be utilized in the fabrication of spin-torque based devices \cite{FinleyAM19}. By calculating the lowest energy configuration, Meinert~{\it et al.} also demonstrated that this alloy is stable as compared to their elemental constituents but not stable with respect to their binary phases thus it is easily decomposes into CoAl and Cr phases~\cite{Meinert13}. On the other hand, it was reported that due to the negative formation energy, Cr$_2$CoAl can possibly stabilize in the inverse (X-type) Heusler structure~\cite{Singh14}. 

In this paper, we report the structural and magnetic properties of single phase Cr$_2$Co$_{(1-x)}$Cr$_x$Al ($x = $ 0, 0.2, 0.4) inverse Heusler alloys. The bulk compositions have been confirmed by energy dispersive x-ray spectroscopy measurements. The Rietveld refinement of XRD data  space group F$\bar{4}$3m confirms the X-type Heusler structure. More interestingly, our analysis with space group I$\bar{4}$m2 reveals a tetragonal distortion in these samples. Interestingly, the vibration of Raman modes with different composition have been observed. The active Raman mode F$_{2g}$ is found to be at about 320~cm$^{-1}$, which confirms the X-type Heusler structure as the A2 and B2 type structures are known to be not Raman active. The area of F$_{2g}$ mode decreases with Cr concentration, which indicate the origin of this mode due to Co vibrations. The magnetic analysis confirms zero moment confirming fully compensated ferrimagnetic nature and complex interactions at low temperatures.


The pollycrystalline samples of Cr$_2$Co$_{(1-x)}$Cr$_x$Al ($x = $ 0, 0.2, 0.4) were prepared by arc melting from CENTORR Vacuum Industries, USA. In order to reduce the oxygen partial pressure, the chamber was evacuated and flushed with high purity argon. This process was repeated few times and  a small piece of Ti metal is melted (which act as getter pump of oxygen) before melting the sample. The appropriate quantities of the constituent elements of 99.99\% purity (from Sigma-Aldrich) were arc melted under an inert argon atmosphere. The ingot was flipped and melted 4-5 times to ensure the homogeneity. For further homogenization and larger grain size samples, the ingot materials were wrapped in Mo foils and sealed in evacuated quartz ampules and then annealed at 1173~K for five days in high temperature box furnace from Nabertherm, GmbH, Germany. The samples were finally quenched in ice water to obtain the highest degree of chemically ordered phase. The bulk compositions have been confirmed by energy dispersive x-ray spectroscopy measurements. We use x-ray diffraction (XRD) with Cu K$\alpha$ ($\lambda$ = 1.5406 $\rm\AA$) radiation for structural study and analyzed the XRD data by Rietveld refinement using FULLPROF package where the background was fitted using linear interpolation between the data points. The Raman measurements were carried out at room temperature using a LabRAM HR evolution Horiba spectrometer. A He-Cd laser with 325~nm excitation wavelength, 1200 lines per mm grating and 1~mW laser power was used. The magnetic measurements are performed using VSM mode in a physical properties measurements system (PPMS EVERCOOL-II) from Quantum design.


In X$_2$YZ type full-Heusler alloys, the unit cell consists of four interpenetrating {\it fcc} sublatices with the positions A (0  0  0), B (0.25, 0.25, 0.25), C (0.5, 0.5, 0.5) and D (0.75, 0.75, 0.75). The perfectly ordered $L2_1$-type structure, with space group Fm$\bar{3}$m, consists $X$ atoms at A and C sites, $Y$ atoms at B site and $Z$ atoms at D site. If we see diagonally from any atom then the sequence of the sites is found to be as $X-Y-X-Z$ \cite{Felser16,Graf11}. In $X_2YZ$ compounds, if the atomic number of Y element is higher than the X element (as the case in the present study) an inverse Heusler structure (also called X-type structure with F$\bar{4}$3m space group) is observed \cite{SkaftourosAPLPRB13}. In this structure half of the $X$ atoms are replaced by $Y$ atoms so the position of the atoms changes as $X$ atoms at A and B sites, $Y$ atoms at C site and $Z$ atoms at D sites. In this case the diagonal sequence of atoms changes as $X-X-Y-Z$, i.e., the difference between the position of two $X$ atoms in particular direction is $\frac{1}{2}$ and $\frac{1}{4}$ in regular and inverse Heusler structures, respectively \cite{Felser16,Graf11,SkaftourosAPLPRB13}.   
\begin{figure}[ht]
\includegraphics[width=3.4in]{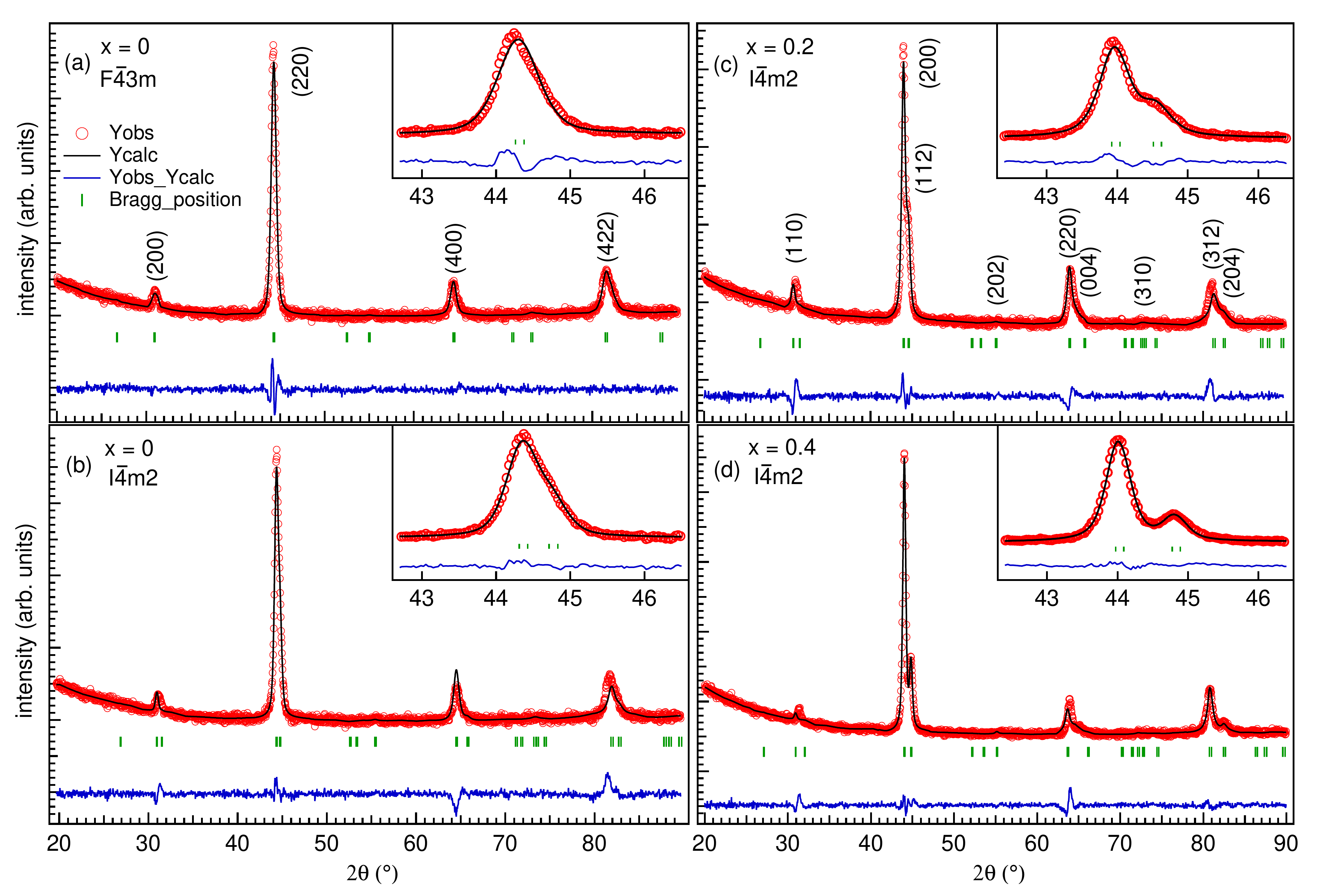}
\caption{The room temperature XRD patterns (a) fitted with space group F$\bar{4}3m$ and (b--d) fitted with space group I$\bar{4}m2$; all the insets are the enlarged view of most intense peak for the sake of clarity of tetragonal nature. XRD patterns (open circles) and Rietveld refinement (black line) of Cr$_2$Co$_{1-x}$Cr$_x$Al ($x=$ 0, 0.2, 0.4) with difference profile (blue line) and Bragg peak positions (short vertical bars, green).}
\label{fig-CCoAl_All_XRD_with_disorder}
\end{figure}
In Figs.~\ref{fig-CCoAl_All_XRD_with_disorder}(c--f), we present the room temperature XRD patterns of Cr$_2$Co$_{1-x}$Cr$_x$Al ($x=$ 0, 0.2, 0.4) recorded in the 2$\theta$ range $20$ to $90^{o}$ and identified to be a single phase. The Rietveld refinement of the XRD pattern with cubic space group F$\bar{4}3m$ (No.~216) confirms the X-type Heusler structure and the obtained lattice constant value is $a=$ 5.784~$\rm \AA$ for the $x=$ 0 sample, as shown in Fig.~\ref{fig-CCoAl_All_XRD_with_disorder}(c), which is consistent with previously reported experimental value ($a=$ 5.794~$\rm \AA$) in ref.~\cite{Jamer_JMMM15}. On the other hand, the calculated value of lattice constant of Cr$_2$CoAl is 5.72~$\rm \AA$ as reported in ref.~\cite{MohantaJMMM17}. We find the fitting reasonably good; however, an asymmetry/distortion towards higher 2$\rm \theta$ is clearly visible in a zoomed view of the 220 reflection [see inset of Fig.~\ref{fig-CCoAl_All_XRD_with_disorder}(c)], which indicates the possibility of further improvement in the fitting.

Therefore, we have again fitted the XRD pattern of the $x=$ 0 sample with the space group I$\bar{4}m2$ (No. 119) i.e. the inverse tetragonal Heusler structure, which found to be a better fit, see Fig.~\ref{fig-CCoAl_All_XRD_with_disorder}(d) and the inset therein. Moreover, for the $x=$ 0.2 and 0.4 samples the XRD patterns, fitted with the space group I$\bar{4}m2$, are shown in the Figs.~\ref{fig-CCoAl_All_XRD_with_disorder}(e, f) and the respective insets are the enlarged view of splitting of the most intense 220 reflection. The Wyckoff position of the atoms of Cr$_2$CoAl alloy in the inverse tetragonal structure; Cr atoms occupy two different sites as Cr1 at 2b (0, 0, 0.5) and Cr2 at 2c (0, 0.5, 0.25). The Co and Al atoms occupy at site 2d (0, 0.5, 0.75) and 2a (0, 0, 0) respectively \cite{Jin_atom_position}.  
 \begin{table}[h]
	\label{tab: The extracted lattice parameters}
	\caption{The lattice parameters of Cr$_2$Co$_{(1-x)}$Cr$_x$Al in the tetragonal phase with I$\bar{4}m2$ space group.}
	\begin{tabular}{p{1.5cm}p{1.5cm}p{1.5cm}p{1.5cm}}
		\hline
		\hline
		$x$$\to$ &  0 &  0.2 &  0.4\\
		$a$ (\rm \AA) & 4.085 & 4.119& 4.169 \\
		$c$ (\rm \AA) & 5.679 &5.682 & 5.698 \\
		$c/a$  & 1.39 &1.38 & 1.37 \\
		\hline  
	\end{tabular}
\end{table}
In order to find a best fit, we had to include the disorder between Cr and Al \cite{LuoPB15,DekaJMMM16}, which found to be about Cr1/Al (55\%/45\%) and Cr2/Al (45\%/55\%) at 2b and 2a sites, respectively for all the samples in inverse tetragonal phase. The lattice constant values obtained from the XRD patterns and tetragonality ratio ($c$/$a$) are summarized in Table~I for all three samples, which agree with the reported values for Cr$_2$CoGa in ref.~\cite{LuoPB15}. We found an increment in the lattice parameters with the substitution of Cr at Co site, which is expected as the atomic radius of the Cr (128~pm) is larger than that of the Co (125~pm). 

In Fig.~2, we show the Raman spectroscopy data of Cr$_2$Co$_{(1-x)}$Cr$_x$Al ($x = $ 0, 0.2, 0.4) measured with 325~nm excitation wavelength at room temperature. In general, it is not easy to observe the Raman signal from the metallic samples; however, it is interesting to note that Zayak {\it et al.} predicted about the experimental evidence of Raman vibrational modes in metallic Heusler alloys \cite{ZayakRaman_PRB05}. 
\begin{figure}[h]
\includegraphics[width=3.4in]{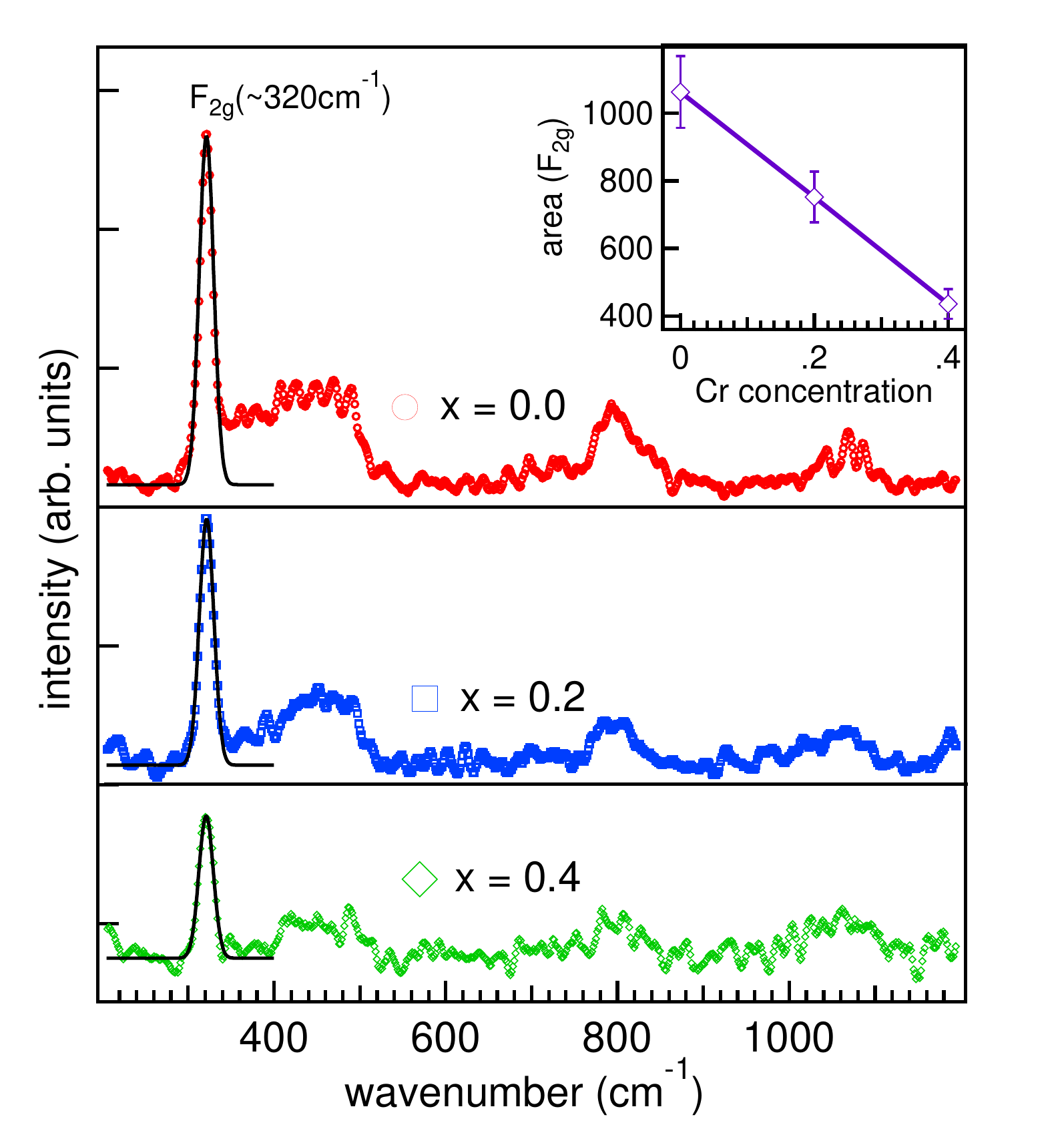}
\caption{Raman spectra of Cr$_2$Co$_{(1-x)}$Cr$_x$Al ($x=$ 0, 0.2, 0.4) measured at room temperature with 325~nm excitation wavelength. Inset shows the area of F$_{2g}$ mode (calculated by fitting the peaks with Gaussian function) with Cr concentration.}
\label{fig2}
\end{figure}
The authors show that for the Heusler alloys, the optical modes are splitted into three well separated triply degenerate modes (F$_{\rm 2g}$+2F$_{\rm 1u}$) where F$_{\rm 2g}$ has been found to be Raman active, whereas two F$_{1u}$ modes are IR active \cite{ZayakRaman_PRB05}. Interestingly, we observed the most intense Raman mode ($F_{\rm 2g}$) at wavenumber $\approx$320~cm$^{-1}$, which is in agreement with the value reported in the literature \cite{Zhan_Raman_12, Zhai_Raman}. It should be noted that Zhan {\it et al.} used Raman scattering to study Co$_2$FeAl Heusler compound across the Curie temperature and observed that the intensity of Raman signal strongly depends on Co atoms vibrational modes. In the present study, we found that the intensity of $F_{\rm 2g}$ peak is decreasing with decrease in the amount of Co, which indicates the origin of this mode due to the vibrations of Co atoms in the lattice. We have fitted the $F_{\rm 2g}$ peak for all three samples with Gaussian function and then plotted the area as a function of Cr concentration, as shown in the inset of Fig.~2. This clearly indicates the decrease in the Raman signal with Cr concentration. Also according to the literature \cite{ZayakRaman_PRB05, Zhan_Raman_12} the existence of Raman modes indicates the presence of X-type structure because A2 and B2 type structures are not Raman active. We have also observed two broad modes at around 800~cm$^{-1}$ and 1025~cm$^{-1}$, which are consistent with reported in ref.~\cite{NehlaJALCOM19} on the similar Heusler alloys.

Further, in order to understand the magnetic properties and moment of Cr$_2$CoAl we have measured isothermal magnetization versus magnetic field (M--H) and magnetic susceptibility $\chi$ versus temperature (M--T). Figs.~3(a, b) show the M--H curves of Cr$_2$Co$_{(1-x)}$Cr$_x$Al ($x=$ 0, 0.4) measured at 300~K and 50~K, respectively, which clearly indicate a very small moment of the order of 10$^{-3}$~$\mu_B/f.u.$ and non saturating behavior of magnetization upto $\pm$ 50~kOe for both the samples. It has been reported that the formation of impure CoAl phase, which is paramagnetic in nature, can be responsible for the linear behavior in the M--H curves \cite{Meinert13}. However, the absence of this CoAl phase in the XRD data of our samples rule out this possibility. Our results suggest an antiferomagnetic coupling between the Cr--Cr neighboring atoms, consistent with ref.~\cite{Singh14}. 
\begin{figure}[h]
\includegraphics[width=3.4in]{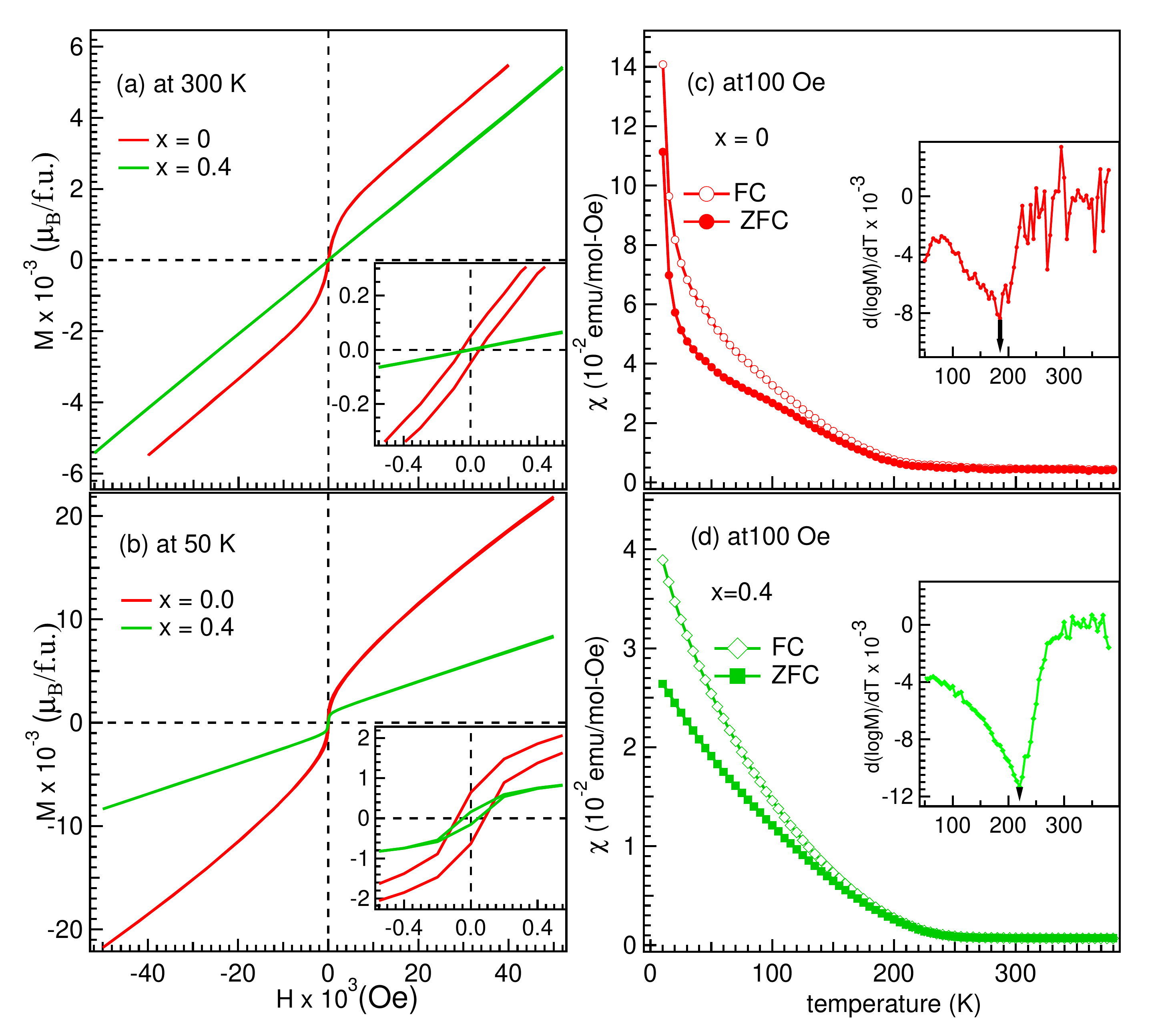}
\caption{Isothermal magnetization $M$ of Cr$_2$Co$_{(1-x)}$Cr$_x$Al ($x=$ 0, 0.4) smaples as a function of applied magnetic field $H$ measured at 300~K (a) and 50~K (b) where respective insets show zoomed view between the range from -0.55~kOe to +0.55~kOe near origin. (c, d) temperature dependent zero-field-cooled (ZFC) and field-cooled (FC) magnetization data of the $x=$ 0 and 0.4 samples, respectively, measured at magnetic field of 100~Oe. Insets in (c, d) present the FC derivative to make the compensation temperature clearly visible.}
\label{fig4}
\end{figure}
On the other hand, at 300~K, a small finite hysteresis has been observed for the $x=$ 0 sample (50~Oe), but not for the $x=$ 0.4 sample, as clearly seen in the zoomed view in the inset of Fig.~3(a). Also, at 50~K, the values of coercivity were found to be 200~Oe and 100~Oe for the $x=$ 0 and 0.4 samples, respectively [inset of Fig.~3(b)], which suggests presence of ferromagnetic order. It has been shown that the partial density of states for minority (majority) spins are located above (below) the Fermi level for Cr1 whereas for Cr2 it is reversed, which implies their spin moments in antiparallel configuration \cite{LuoPB15}. The observed magnetic behavior in Figs.~3(a, b) is consistent as the Co at Y site couples ferromagnetically with the nearest neighbor Cr and antiferromagnetically with second nearest neighbor Cr and the two distinct Cr sites couple ferrimagnetically \cite{MohantaJMMM17}. This type of alignment is attributed due to a competition between the intra-atomic exchange splitting of the magnetic atom $d$ states and the inter-atomic covalent interaction of $d$ states from  atoms at different sites \cite{KublerPRB83}. This also agrees with the theoretical predication of half-metallic fully compensated ferrimagnetic nature in the Cr$_2$CoAl sample \cite{Meinert13, Jin_atom_position}. 

In Figs.~3(c, d), we show the magnetic susceptibility $\chi$ as a function of temperature in both zero-field-cooled (ZFC) and field-cooled (FC) modes for the $x=$ 0 and 0.4 samples, respectively measured in the temperature range of 10--380~K and at 100~Oe magnetic field. For the $x=$ 0 sample, the moment is fairly constant and close to zero, i.e., the sample is in fully compensated ferrimagnetic state above 200~K, which is defined a compensation temperature, as marked in the derivative shown in the inset of Fig.~3(c). Interestingly, the moment increases significantly below $\approx$ 200~K and the irreversible behavior between FC and ZFC moments is clearly evident. Furthermore, the ZFC curve shows a down hump $\approx$ 50~K and below around 25~K there is a sharp increase in the moment for both FC and ZFC modes. The behavior of temperature dependent magnetization for the $x=$ 0 sample is in good agreement with the only reported in ref.~\cite{Jamer_JMMM15}. On the other hand, it is important to note here that Jamer {\it et al.} attributed the low temperature behavior as paramagnetic due to the presence of extra CoAl phase in their sample \cite{Jamer_JMMM15}, which clearly is not the case in the present study as evident in our XRD analysis shown in Fig.~1. With increasing Cr concentration at Co site, i.e., for the $x=$ 0.4 sample in Fig.~3(d), the compensation temperature shifted to around 220~K, which is clearly visible in the derivative plot as shown in the inset. Below this temperature, we observe bifurcation in ZFC and FC curves and consistently increase in the moment till lowest measured temperature. This magnetization behavior for both the samples suggests for a complex magnetic interactions at low temperatures, which motivates to study these materials using neutron diffraction for further understanding these interactions and the role/possibility of atomic disorder \cite{NehlaPRB19}. Though the Curie temperature (T$_{\rm C}$) is expected at around 700~K \cite{Jamer_JMMM15}, the magnetization measurement at high temperature would require to find exact value of T$_{\rm C}$ of these materials \cite{MT_Midhunlal}. Interestingly, the nature of competing magnetic behavior is clear from the irreversibility in ZFC--FC curves where the fully compensated feature is very sensitive to the compositions \cite{StinshoffPRB17}. 

Note that the half-metallic full-Heusler alloys follow a Slater-Pauling behavior, i.e., the total spin magnetic moment per formula unit, M$_t$, in $\mu_B$ scales with the total number of valence electrons, Z$_t$, following the rule: M$_t$ = Z$_t$--24 \cite{SkaftourosAPLPRB13}. In the case of fully compensated ferrimagnet (FCF), the total spin magnetic moment should be zero like Cr$_2$CoGa and Fe$_2$VGa having exactly 24 valence electrons and the ground state of Cr$_2$CoAl is found to be ferrimagnetic in inverse Heusler structure \cite{Meinert13}. However, the ferrimagnetic state of the Cr$_2$CoAl sample arises due to the antiferromagnetic coupling of Cr-Cr atoms of inequivalent nearest neighbors and show the nature of fully compensated ferrimagnetic \cite{Meinert13}. In a more precise way, for these type of alloys, there is a competition between the magnetic states of atoms, which decides whether the alignment of the moments should be ferromagnetic or antiferromagnetic. Therefore, Cr$_2$CoAl possess almost vanishing total spin magnetic moments in FCF states because of the antiferromagnetic alignment of the Cr-Cr inequivalent nearest neighbouring atoms due to the direct interaction between $d$ states \cite{Singh14}. Moreover, the Curie temperature of these alloys is expected to be high \cite{Galanakis11}, which makes them most promising candidate for devices.


In summery, we successfully prepared single phase Cr$_2$Co$_{(1-x)}$Cr$_x$Al ($x=$ 0, 0.2, 0.4) and investigated the structural and magnetic behavior using x-ray diffraction (XRD), Raman spectroscopy, isothermal magnetization, and magnetic dc susceptibility measurements. The Rietveld refinement of XRD patterns confirm the X-type Heusler structure, and interestingly a tetragonal distortion (space group I$\bar{4}m2$) has been observed in these samples where the c/a value is found to be around 1.38. The active Raman mode F$_{2g}$ is found to be at about 320~cm$^{-1}$, which confirms the X-type Heusler structure as the A2 and B2 type structures are known to be not Raman active. The area of F$_{2g}$ mode decreases with Cr concentration, which suggests the origin of this mode due to Co vibrations. The susceptibility data show irreversible behavior between ZFC and FC curves, which indicate complex magnetic interactions. The magnetization data show that the magnetic moment is close to zero at around 70~kOe, which confirms the fully compensated ferrimagnetic nature of these inverse Heusler alloys.


This work was financially supported by the BRNS through DAE Young Scientist Research Award to RSD with project sanction No. 34/20/12/2015/BRNS. MS, GDG and PN acknowledge the MHRD, India for fellowship through IIT Delhi. Authors acknowledge various experimental facilities at IIT Delhi like XRD and PPMS EVERCOOL-II at Physics department; glass blowing section, SEM and EDX at central research facility (CRF); Raman spectroscopy at nano research facility (NRF).\\


\end{document}